\documentclass[11pt,twoside]{article}
\usepackage{asp2014}

\aspSuppressVolSlug
\resetcounters

\begin{document}

\title{An update on the development of ASPIRED}

% full name: Marco Lam
\author{Marco~C.~Lam$^{1,2}$, Robert~J.~Smith$^1$, Josh~Veitch-Michaelis$^{1,3}$, Iain~A.~Steele$^1$, and Paul R. McWhirter$^4$}
\affil{$^1$Astrophysics Research Institute, Liverpool John Moores University, IC2, LSP, 146 Brownlow Hill, Liverpool L3 5RF, UK \email{c.y.lam@ljmu.ac.uk}}
\affil{$^2$School of Physics and Astronomy, Tel Aviv University, Tel Aviv 69978, Israel}
\affil{$^3$IceCube Particle Astrophysics Center, Department of Physics, University of Wisconsin-Madison, Madison, WI 53706, USA}
\affil{$^4$Instituto de Astrof\'isica de Canarias~(IAC), Calle V\'ia L\'actea s/n, E-38200 La Laguna, Tenerife, Spain}

% remove/add as you need

% remove/add authors as you need
\paperauthor{Marco~C.~Lam}{c.y.lam@ljmu.ac.uk}{0000-0002-9347-2298}{Astrophysics Research Institute}{Liverpool John Moores University}{Liverpool}{Merseyside}{L3 5RF}{UK}
\paperauthor{Marco~C.~Lam}{c.y.lam@ljmu.ac.uk}{0000-0002-9347-2298}{School of Physics and Astronomy}{Tel Aviv University}{Tel Aviv}{}{69978}{Israel}
\paperauthor{Robert~J.~Smith}{r.j.smith@ljmu.ac.uk}{}{Astrophysics Research Institute}{Liverpool John Moores University}{Liverpool}{Merseyside}{L3 5RF}{UK}
\paperauthor{Josh~Veich-Michaelis}{j.veitchmichaelis@ljmu.ac.uk}{0000-0003-2780-7843}{Astrophysics Research Institute}{Liverpool John Moores University}{Liverpool}{Merseyside}{L3 5RF}{UK}
\paperauthor{Josh~Veich-Michaelis}{j.veitchmichaelis@ljmu.ac.uk}{0000-0003-2780-7843}{IceCube Particle Astrophysics Center}{University of Wisconsin-Madison}{Madison}{Wisconsin}{53706}{USA}
\paperauthor{Iain~A.~Steele}{i.a.steele@ljmu.ac.uk}{0000-0001-8397-5759}{Astrophysics Research Institute}{Liverpool John Moores University}{Liverpool}{Merseyside}{L3 5RF}{UK}
\paperauthor{Paul~R.~McWhirter}{ross.mcwhither@iac.es}{0000-0002-6616-0782}{Instituto de Astrof\'isica de Canarias~(IAC)}{Calle V\'ia L\'actea s/n}{E-38200 La Laguna}{Tenerife}{}{Spain}

% remove/add as you need

% leave these next few aindex lines commented for the editors to enable them. Use Aindex.py to generate them for yourself.
% first presenting author should be the first entry for bold-facing the author index page-reference
%\aindex{Lam,~M.}
%\aindex{Author2,~S.}
% remove/add as you need

% leave the ssindex lines commented for the editors to enable them, use Index.py to suggest yours
%\ssindex{FOOBAR!conference!ADASS 2020}
%\ssindex{FOOBAR!organisations!ASP}

% leave the ooindex lines commented for the editors to enable them, use ascl.py to suggest yours
%\ooindex{FOOBAR, ascl:1101.010}
  
\begin{abstract}
We are reporting the updates in version 0.2.0 of the Automated SpectroPhotometric REDuction (\textsc{ASPIRED}) pipeline, designed for common use on different instruments. The default settings support many typical long-slit spectrometer configurations, whilst it also offers a flexible set of functions for users to refine and tailor-make their automated pipelines to an instrument's individual characteristics. Such automation provides near real-time data reduction to allow adaptive observing strategies, which is particularly important in the Time Domain Astronomy. Over the course of last year, significant improvement was made in the internal data handling as well as data I/O, accuracy and repeatability in the wavelength calibration.
\end{abstract}

\section{Introduction}
\textsc{ASPIRED} - Automated SpectroPhotometric REDuction~(\citealt{2019arXiv191205885L}, hereinafter LSS19; \citealt{marco_c_lam_2020_4306065}), is a spectral data reduction software, written in \textsc{Python 3}, that is designed for common use on different instruments. Over the course of last year, it has been tested on several active instruments: SPRAT on the Liverpool Telescope~(LT, \citealt{LT}), ISIS on the William Herschel Telescope\footnote{\url{https://github.com/cylammarco/ASPIRED-example}}~\citep{2020MNRAS.493.6001L}, GMOS long slit~(LS) mode on the Gemini Telescopes North~(GTN)\footnote{\url{https://github.com/cylammarco/bhtomspec}}, and OSIRIS LS mode on the Gran Telescopio Canarias~(GTC)\footnote{\url{https://github.com/P-R-McWhirter/pyOsiris}}. Apart from LT/SPRAT, the test efforts are \textbf{not} in collaboration with the respective observatories.

\section{What's new?}
In the latest release, we have abstracted the spectral handling into a \verb+_spectrum1D+ class to handle all the data I/O of the extracted spectra. Wavelength and flux calibrations are operating on these objects, each level of calibration is stored and can be recalled for diagnostics. As laid down in the design requirements (LSS19), the spectral data reduction is modularised into 3 components at the highest level. We have a set of API that maps to every function available with the wavelength calibration software, \textsc{RASCAL}~\citep{2019arXiv191205883V, veitch_michaelis_joshua_2020_4117517}. They are both available through PyPI~(i.e. \verb+pip install aspired rascal+). Documentations are frequently updated on Readthedocs\footnote{\url{https://aspired.readthedocs.io/}}. Files can be exported as FITS or CSV. More effort is still required to allow manual input or changes of FITS header keywords or comments. Automated testing is moved from Travis CI to GitHub Actions to simplify the process -- unit tests and build tests are performed with Python 3.6, 3.7 and 3.8 on the latest versions of Windows, Ubuntu and MacOS~(see more at \url{https://github.com/features/actions}).

\section{Application and Comparison with the current SPRAT pipeline}

The red mode of LT/SPRAT with a Xe arc is our primary testing system and the data have most of the features and capabilities expected of an observatory's common-user, long-slit spectrograph. We are currently commissioning the new automated pipeline for it. This will be further extended to deploy a common pipeline for both LT/SPRAT and the MOOKODI spectrograph on the Lesedi Telescope in 2021. A separate prototype pipeline was implemented using YAML file as configuration file to fully control data reduction for refining LT/SPRAT and extracting GTN/GMOS LS spectra.

\begin{figure}
    \centering
    \includegraphics[width=\textwidth]{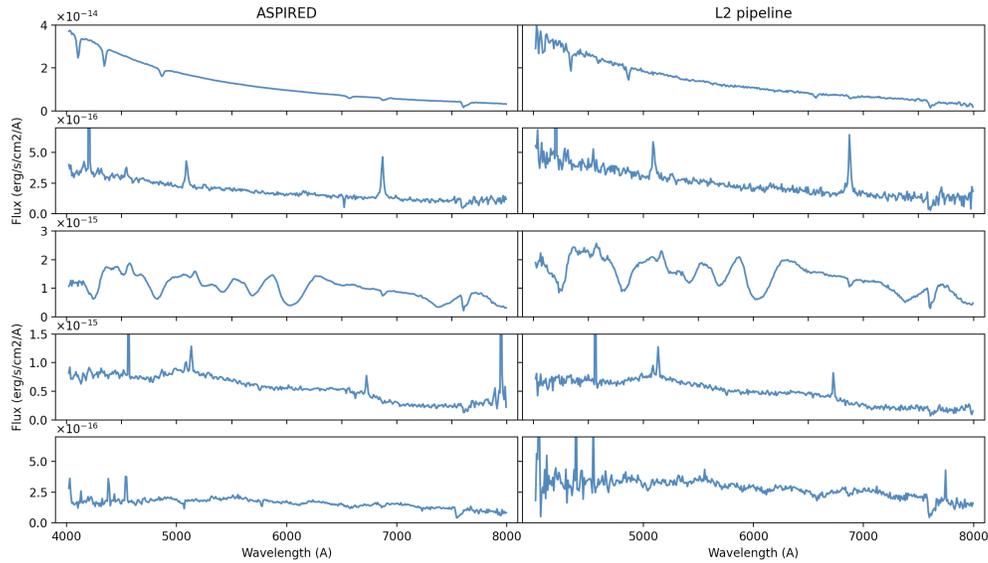}
    \caption{Some examples comparing results from the two pipelines are shown in accompanying figure. In virtually all cases the \textsc{ASPIRED} extraction is obviously higher in the signal-to-noise.
}
    \label{fig:comparison}
\end{figure}

\begin{figure}
    \centering
    \includegraphics[width=\textwidth]{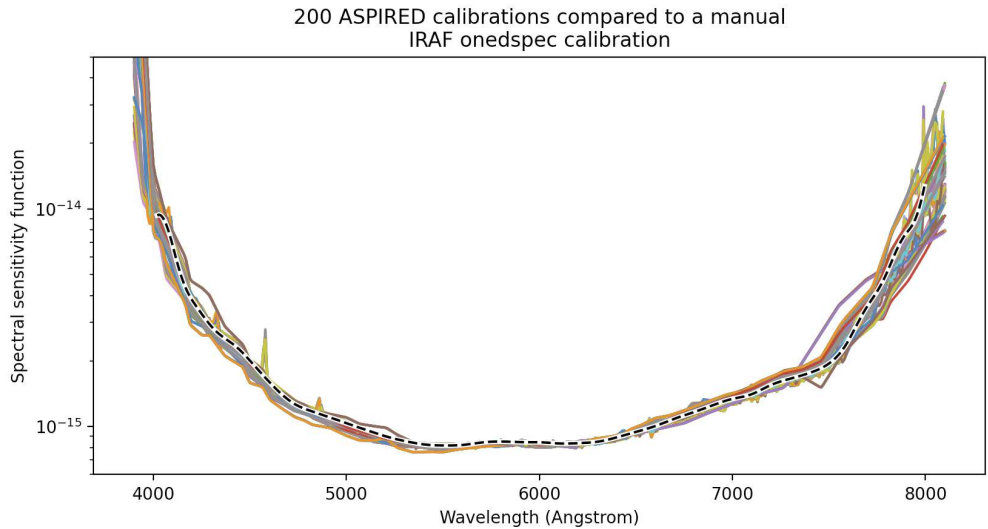}
    \caption{In an alternative check we compare the automated spectral sensitivity curves generated by \textsc{ASPIRED} on default settings (coloured lines) to a calibration that was manually and interactively derived in IRAF by an experienced observer combining observations of multiple different spectrophotometric standards (dashed line). They are all normalised to each other to compare just the shapes.
}
    \label{fig:sensitivity}
\end{figure}

As one element of testing of the new software we compare its results to those from known and established software packages. Comparisons are based on a complete sample of all data taken with the LT/SPRAT~\citep{2014SPIE.9147E..8HP} spectrometer over two months~(Jan - Feb 2020). See some examples in Figure~\ref{fig:comparison} \& \ref{fig:sensitivity}.
The sample includes 493 spectra with both point sources and extended targets in all observing conditions, from excellent to thick cloud or bright twilight, i.e., this tests all qualities of data a user is likely to encounter, good and bad.
We compare performance of the new \textsc{ASPIRED} pipeline to the existing SPRAT pipeline (`L2 pipeline') that was custom developed alongside the instrument and optimised to both the hardware and LT operational model. The full sample was passed through both pipelines. All raw data and output spectra were qualitatively assessed and rated as to whether the extracted spectrum was a reasonable representation of the data:
\begin{itemize}
    \item Both \textsc{ASPIRED} and L2 produce believable result 401 (81\%)
    \item Unusable data: cloud/observer error/technical faults 38 (8\%)
    \item \textsc{ASPIRED} unable to extract a spectrum, but L2 pipeline did 52 (11\%)
    \item L2 pipeline unable to extract a spectrum, but \textsc{ASPIRED} did 1 (0\%)
    \item Neither pipeline extracted a spectrum 1 (0\%)
\end{itemize}

The vast majority of the \textsc{ASPIRED} `failures' are extremely low signal-to-noise spectra where the bespoke L2 pipeline was able to use forced extraction through its a priori knowledge of the instrument's configuration. In contrast \textsc{ASPIRED} was being run using generic defaults that require it to detect the source in the slit. It successfully extracted 91\% of the frames that the bespoke pipeline did and achieved higher signal-to-noise (Figure~\ref{fig:comparison}). We do not optimise \textsc{ASPIRED} defaults to the one SPRAT instrument. A user would be able to provide such instrument-specific knowledge into any pipeline.

\section{Wavelength Calibration - \textsc{RASCAL}}

\textsc{RASCAL} -- RANSAC-Assisted Spectral Calibration is designed to solve for the pixel-wavelength function requiring only an atlas of calibration lines (wavelength), a list of peaks (peaks), and some information about the system (wavelength range).
The development in the past year has (1) improved the robustness; (2) added more instrumental systems; (3) a cleaned line list that is more suited for low/medium resolution spectrograph; and (4) allowed users to supply custom lines. These improvements allow rapid convergence of the solution when fitted with a fine tuned line list provided by the observatory.
A number of examples are available at Readthedocs page\footnote{\url{https://rascal.readthedocs.io/en/latest/}}, including: (1)~Liverpool Telescope - SPRAT (red mode), (2)~William Herschel Telescope - ISIS (R300R), (3)~New Technology Telescope - EFOSC (Gr\#11), (4)~Gemini North - GMOS-LS (R400+G5305), (5)~Gran Telescopio Canarias OSIRIS-LS (R1000B), (6)~Issac Newton Telescope - IDS (R1200U), and (7)~Keck II Telescope - DEIMOS (830G).

\section{Development and Release Timeline}

\textsc{ASPIRED} is currently at version v0.2.0, we expect updates and bug fixes in month-scale. We will continue to expand and improve the API to provide more functions to delivery higher quality and flexibility. The concurrent development of \textsc{RASCAL} will continue to support \textsc{ASPIRED} and we are looking to expand to handle calibrations for different variants of multi-object spectrograph.

% For example in \citet{PID_adassxxx} it was shown that ...

\bibliography{P10-241}

% if we have space left, we might add a conference photograph here. Leave commented for now.
% \bookpartphoto[width=1.0\textwidth]{foobar.eps}{FooBar Photo (Photo: Any Photographer)}

\end{document}